\documentclass[prl,twocolumn,aps,superscriptaddress,longbibliography]{revtex4-2}
\usepackage{graphicx}
\usepackage[fleqn]{amsmath}
\usepackage{amssymb}
\usepackage[dvipsnames]{xcolor}
\usepackage[normalem]{ulem}
\definecolor{goodgreen}{rgb}{0.1,0.5,0}
\definecolor{goodred}{rgb}{0.7,0,0}
\usepackage{float}
\usepackage{multirow}
\usepackage[colorlinks,urlcolor=goodgreen,citecolor=blue,linkcolor=goodred]{hyperref}
\usepackage{bm}
\usepackage{physics}
\usepackage{graphicx}
\usepackage{booktabs}
\allowdisplaybreaks
\newcommand{\rem}[1]{}

\begin{document}
\title{Nonlinear Mode Coupling in Silicon Nitride Membrane Resonators}





\author{Soumya Kanti Das$^{\hyperlink{equal}{*}}$}

\affiliation{Department of Physics, \\Indian Institute of Science Education and Research Bhopal, Bhopal-462066, Madhya Pradesh, India}

\author{Nishta Arora$^{\hyperlink{equal}{*}\,\hyperlink{corrA}{\dagger}}$}

\affiliation{Centre for Nano Science and Engineering, \\Indian Institute of Science, Bengaluru-560012, Karnataka, India}



\author{Hridhay A S}
\affiliation{Department of Physics, \\Indian Institute of Science Education and Research Bhopal, Bhopal-462066, Madhya Pradesh, India}

\author{Akshay Naik}
\affiliation{Centre for Nano Science and Engineering, \\Indian Institute of Science, Bengaluru-560012, Karnataka, India}

\author{Chandan Samanta$^{\hyperlink{corrB}{\ddagger}}$}
\affiliation{Department of Physics, \\Indian Institute of Science Education and Research Bhopal, Bhopal-462066, Madhya Pradesh, India}



\begin{abstract}


\textbf{Nonlinear interactions between vibrational modes play a crucial role in understanding the dynamical response of nanomechanical resonators. Here, we report the experimental observation and theoretical modeling of nonlinear mode coupling in a high-stress square silicon nitride membrane resonator. We quantify frequency shifts of the fundamental mode arising from tension-mediated geometric nonlinearity by increasing the amplitude of the fundamental mode and higher-order flexural modes. A quantitative theoretical framework based on Kirchhoff–Love plate theory is developed, which incorporates both intrinsic Duffing nonlinearity and nonlinear intermodal coupling and shows good agreement with experimental measurements for the (1,1)-(2,1) and (1,1)-(2,2) mode pairs. We further compute the nonlinear coupling matrix across mode families, revealing the role of mode symmetry and spatial overlap in governing intermodal interactions. These results establish nonlinear mode coupling as a controllable resource for multimode frequency tuning and mechanical transduction.}

\end{abstract}

\maketitle
Mechanical resonators\cite{bachtold2022mesoscopic, xu2022nanomechanical} play an important role in both fundamental research and emerging technologies. Their exceptional sensitivity to external stimuli
underpins advances across a wide range of research areas, including force sensing\cite{mamin2001sub,gavartin2012hybrid,de2018ultrasensitive}, mass spectrometry\cite{naik2009towards,dominguez2018neutral}, magnetometry\cite{forstner2012cavity, rossi2019magnetic, vsivskins2020magnetic, yousuf2025mechanical}, accelerometers\cite{krause2012high,bawden2025precision}, pressure sensors\cite{reinhardt2024self, green2025accurate,chen2022nano},  magnetic resonance imaging\cite{poggio2010force,degen2009nanoscale,grob2019magnetic}, and scanning probe microscopy\cite{de2017universal,rossi2017vectorial}. Beyond sensing, mechanical resonators are widely used in communications and signal 
processing\cite{mahboob2011interconnect,romero2024acoustically,zhang2025few}. They serve as key interfaces in hybrid quantum systems\cite{kurizki2015quantum, clerk2020hybrid}, coupling mechanical motion to optical\cite{barzanjeh2022optomechanics,chu2020perspective}, electrical\cite{naik2006cooling, urgell2020cooling}, and spin\cite{rugar2004single, thomas2021entanglement, fedele2025coupling} degrees of freedom. These capabilities establish mechanical resonators as a unifying platform bridging classical and quantum physics\cite{teufel2011sideband,wollack2022quantum,samanta2023nonlinear,yang2024mechanical}, combining technological relevance with fundamental discovery\cite{campbell2021searching,manley2024microscale}.

Among various classes of mechanical resonators, high-stress silicon nitride ($Si_3N_4$) membranes have emerged as a leading platform. Their resonance frequencies, quality factor, and mechanical mode shapes can be precisely engineered through device geometry and boundary conditions. Their exceptionally high quality factors\cite{tsaturyan2017ultracoherent, yuan2015silicon,chakram2014dissipation,ji2021methods,engelsen2024ultrahigh,xi2025soft} and compatibility with chip-scale integration make these resonators well suited for precision sensing and for probing mechanical dynamics at room and cryogenic temperatures\cite{purdy2012cavity,gisler2022soft}. As a result, these resonators are playing an increasingly important role in exploring light-matter interaction\cite{thompson2008strong,purdy2013observation}, ground state cooling\cite{rossi2018measurement}, entanglement\cite{chen2020entanglement}, optomechanical squeezing\cite{purdy2013strong}, optical-to-microwave photon transducers\cite{andrews2014bidirectional}, quantum memory\cite{kristensen2024long}, gravitational-wave detection\cite{gely2021superconducting}, dark matter detector\cite{manley2021searching}, and advanced precision sensing\cite{mason2019continuous}.

Most studies of $\mathrm{Si_3N_4}$ membrane resonators have focused on single, isolated vibrational modes \cite{fink2016quantum, snell2022heat, yang2019spatial}, whereas practical membranes support a dense spectrum of mechanical resonances whose interactions can strongly influence device dynamics and performance. At sufficiently large oscillation amplitudes, mechanical resonators enter into a nonlinear regime governed by tension-induced geometric nonlinearity, enabling nonlinear modal coupling that gives rise to amplitude-dependent frequency tuning, energy exchange between orthogonal modes, and dynamic control of the mechanical response\cite{asadi2018nonlinear}. These effects enable multimode functionalities including hybridized resonances\cite{mestre2025network,prasad2019gate}, enhanced energy transfer\cite{li2025cascade,arora2022qualitative}, bandwidth broadening \cite{arora2024mixed}, and tunable dissipation pathways \cite{guttinger2017energy, kecskekler2021tuning,prasad2023tunable}. Despite their importance, nonlinear modal interactions in square $\mathrm{Si_3N_4}$ membranes remain largely unexplored, even though they are crucial for advanced sensing, signal processing, and hybrid classical–quantum transduction architectures.

In this work, we present an experimental and theoretical study of nonlinear intra- and intermodal coupling in a high-stress square $\mathrm{Si_3N_4}$ membrane resonator. By driving multiple flexural modes into the nonlinear regime, we observe amplitude-dependent frequency shifts arising from both intramodal (intrinsic Duffing nonlinearity) and intermodal coupling. We develop a simple yet quantitative theoretical framework based on Kirchhoff–Love plate theory that accurately captures these effects. Through a systematic mapping of coupling strengths across mode families, we compute the coupling matrix between the fundamental and higher-order modes, revealing the role of modal overlap in governing nonlinear interactions. Furthermore, we compute the selective frequency tuning of a target mode via the excitation of orthogonal modes, establishing nonlinear mode coupling as a powerful resource for engineering multimode functionality in high-stress $\mathrm{Si_3N_4}$ membrane resonators.
%
%
\begin{figure}
    \centering
    \includegraphics[width=1\linewidth]{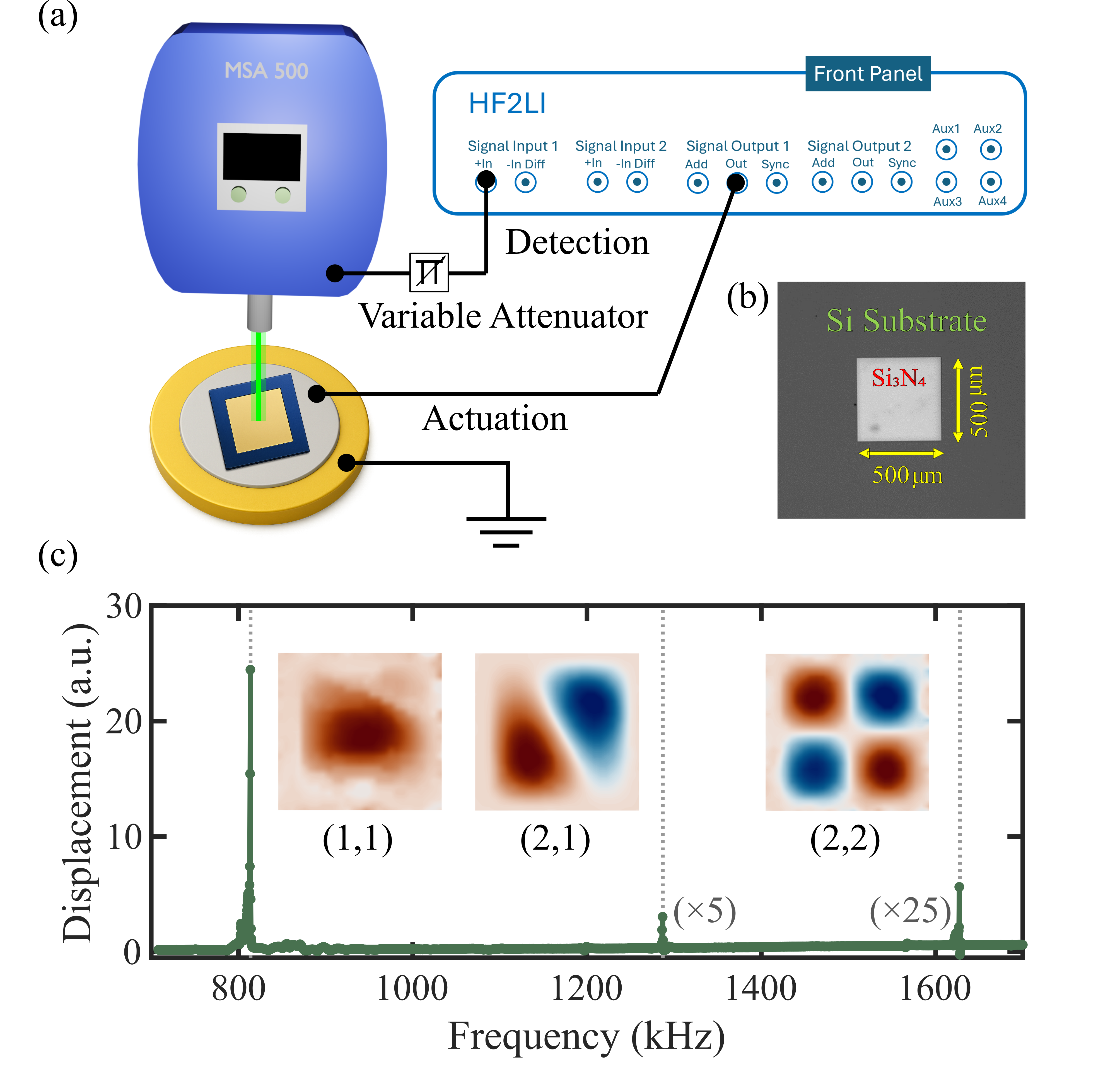}
    \caption{ Measurement setup and vibrational modes of the device. (a) Schematic of the laser Doppler vibrometry (LDV) based measurement setup. A Polytec MSA-500 system is used to optically probe the out-of-plane motion of $\mathrm{Si_3N_4}$ membrane, while electrical actuation and signal detection are performed using a Zurich instruments HF2LI lock-in amplifier. (b) Optical micrograph of the device, showing a $500\,\mu\mathrm{m} \times 500\,\mu\mathrm{m}\ \mathrm{Si_3N_4}$ membrane suspended over a silicon substrate. (c) Measured displacement amplitude spectrum with drive frequency showing multiple mechanical resonances. Higher frequency modes are scaled by a factor of $\times 5$ and $\times 25$ for visibility. Inset shows the corresponding spatial mode shapes of the resonant modes.}
    \label{fig:Figure 1}
\end{figure}

The experiments are performed on a suspended silicon nitride square membrane from Norcada with side length $(L)$ of 500 $\mu$m and a thickness $(h)$ of 100 nm clamped to a 500 $\mu$m thick silicon substrate. The $Si_3N_4$/Si substrate is mounted on a piezoelectric actuator that enables out-of-plane excitation via an applied AC voltage $V(t) = V_0 \cos(\omega t)$. The frequency-amplitude response of the membrane is measured optically using a laser Doppler vibrometer (Polytec MSA 500). All measurements were performed under high vacuum at a base pressure of $2 \times 10^{-5}\,\mathrm{mbar}$ to minimize air damping. Figure \ref{fig:Figure 1}(a) shows a schematic of the experimental setup, in which a lock-in amplifier (Zurich Instruments HF2LI) is used to simultaneously actuate and detect the mechanical motion of the membrane. An optical micrograph of the membrane and the measured mode spectrum between 700 and 1700 kHz are presented in Figs.~\ref{fig:Figure 1}(b) and \ref{fig:Figure 1}(c), respectively. Three vibrational modes are observed within this frequency range; the inset shows the measured spatial mode shapes, with modes labeled $(n,m)$ according to the number of anti-nodes along the two in-plane directions of the membrane.

\begin{figure}
    \centering
    \includegraphics[width=1\linewidth]{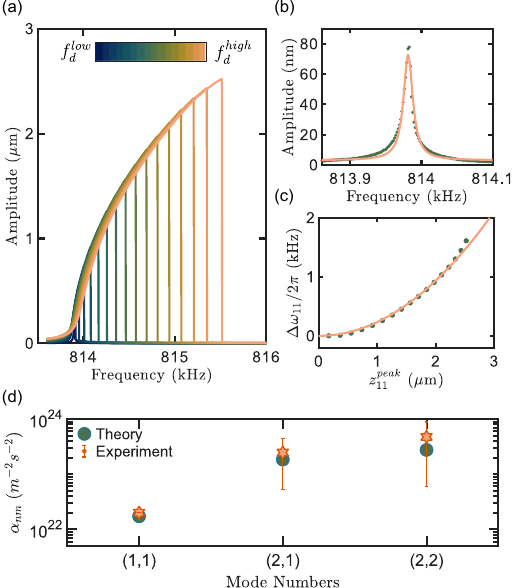}
    \caption{Intramodal coupling and Duffing constant. (a) Nonlinear frequency response of the $(1,1)$ mode, measured at increasing drive strengths. (b) Linear response of the $(1,1)$ mode at the lowest drive amplitude; solid lines indicate Lorentzian fits used to extract the resonance frequency and quality factor. (c) Amplitude-dependent frequency shift corresponding to the maximum displacement amplitudes extracted from (a). Solid line show fit using Eq. (\ref{6}) to extract the Duffing coefficient. (d) Comparison of the experimentally extracted Duffing coefficients with theoretical values calculated from Eq. (\ref{5}) for the $(1,1)$, $(2,1)$, and $(2,2)$ modes.}
    \label{fig:figure 2} 
\end{figure}

We first characterize the out-of-plane vibrational modes at low drive amplitudes, where both intramodal and intermodal coupling are negligible. Figure \ref{fig:figure 2}(b) shows the linear frequency response of the $(1,1)$  mode, while the responses of the $(2,1)$ and $(2,2)$ modes are presented in Fig. S4. In this regime, all modes exhibit Lorentzian line shapes. The measured resonance frequencies of the $(1,1)$, $(2,1)$, and $(2,2)$ modes are 814 kHz, 1286 kHz, and 1628 kHz, with corresponding quality factors of $5.2\times10^{4}$, $12.5\times10^{4}$, and $11.3\times10^{4}$, respectively. 
Figure~\ref{fig:figure 2}(a) displays the nonlinear frequency response of the $(1,1)$ mode, measured at increasing drive amplitudes, arising from intrinsic Duffing nonlinearity.

We describe the membrane dynamics within the framework of Kirchhoff–Love plate theory\cite{leissa1969vibration}. The transverse displacement $w(x,y,t)$ satisfies 
\begin{equation}
\begin{aligned}
    D \nabla^4 w(x, y, t) 
    &- \left( N_{xx} \frac{\partial^2 w}{\partial x^2} 
    + 2N_{xy} \frac{\partial^2 w}{\partial x \partial y} 
    + N_{yy} \frac{\partial^2 w}{\partial y^2} \right) \\
    &\quad + \rho h \frac{\partial^2 w}{\partial t^2} 
    = 0,
   \label{1}
\end{aligned}
\end{equation}
%
where $D$ is the flexural rigidity, $N_{xx},  N_{xy}, N_{yy}$ are the in-plane stress resultants, $\rho$ is the mass density of the membrane and $h$ is the thickness. 
%
%
Within the high stressed thin membrane approximation\cite{vogl2006nonlinear} the in-plane stress components can be expressed as
\begin{equation}
    \begin{aligned}
    N_{xx} &= Eh \left( \epsilon_0 + \frac{1}{2} \left( \frac{\partial w}{\partial x} \right)^2 \right), \\
    N_{yy} &= Eh \left( \epsilon_0 + \frac{1}{2} \left( \frac{\partial w}{\partial y} \right)^2 \right), \\
    N_{xy} &= \frac{1}{2} Eh \left( \frac{\partial w}{\partial x} \frac{\partial w}{\partial y} \right),
    \end{aligned}
    \label{2}
\end{equation}
where $E$ is the Young's modulus, and $\epsilon_0$ is the built-in tensile strain of the membrane. The transverse displacement is expressed as $w(x,y,t) = \phi(x,y)\, z(t)$, where $\phi(x,y) = \sin\!\left(\frac{n\pi x}{L}\right)\sin\!\left(\frac{m\pi y}{L}\right)$
is the spatial mode shape and \( z(t) \) is the corresponding out of plane displacement.
Substituting the mode expansion into Eq. \eqref{1} and applying the Galerkin discretizations procedure, followed by the inclusion of linear damping ($\nu$) and external driving force ($f_d \, cos(\omega_d t)$), yields the Duffing equation of motion
\begin{equation}
    \ddot{z}\,+ \omega_{nm}^2 z\,+ \nu \dot{z} \, + \alpha_{nm} z^3\, = f_d \, cos(\omega_d t),
    \label{3}
\end{equation}
The expressions for the linear resonance frequency $\omega_{nm}$ and the Duffing coefficient $\alpha_{nm}$ for a general mode $(n,m)$ are given by
\begin{equation}
    \begin{aligned}
    \omega_{nm}^2 & = \frac{E\epsilon_0\pi^2 (n^2+m^2)}{\rho  L^2},
    \label{4}
    \end{aligned}
\end{equation}
\begin{equation}
    \begin{aligned}
    \alpha_{nm} & = \frac{3E}{32\rho}\left(\frac{\pi}{L}\right)^4\left[n^4+m^4-\frac{2}{3}n^2m^2\right].
    \label{5}
    \end{aligned}
\end{equation}
%

We use the measured resonance frequency of the fundamental $(1,1)$ mode to extract the built-in tensile strain $\epsilon_{0} = 3.7 \times 10^{-3}$ from Eq. (\ref{4}). This value is subsequently used to estimate the theoretical resonance frequencies of higher-order modes, shown by the dashed lines in Fig. \ref{fig:Figure 1}(c), which are in good agreement with the experimentally observed higher mode frequencies. Solving Eq. \eqref{3} using the method of multiple scales yields an amplitude-dependent nonlinear pulling of the resonance frequency,
\begin{equation}
    \omega_{nm}^{peak} = \omega_{nm} \, +\, \frac{3\alpha_{nm}}{8\omega_{nm}}(z^{peak})^2,
    \label{6}
\end{equation}
where $z^{peak}$ denotes the maximum displacement amplitude of the $(n,m)$ mode. The measured values of $z^{peak}$  is plotted as a function of the shift in peak frequency $(\omega_{nm}^{peak}-\omega_{nm})$ for the $(1,1)$ mode in Fig. \ref{fig:figure 2}(c). The data is fitted using Eq. (\ref{6}), with Duffing constant as the sole free parameter. The experimental Duffing coefficients for the $(1,1)$, $(2,1)$, and $(2,2)$ modes are extracted and found to be $2\times10^{22}$ $m^{-2}s^{-2}$, $2.5\times10^{23}$ $m^{-2}s^{-2}$, and $4.7\times 10^{23}$ $m^{-2}s^{-2}$, respectively. Figure \ref{fig:figure 2}(d) shows good agreement between the experimentally obtained Duffing coefficients and the theoretical values calculated from Eq. (\ref{5}). The corresponding analysis for the $(2,1)$ and $(2,2)$ modes are presented in Fig.~S4 of the Supplementary Material.


%
\begin{figure}
    \centering
    \includegraphics[width=1\linewidth]{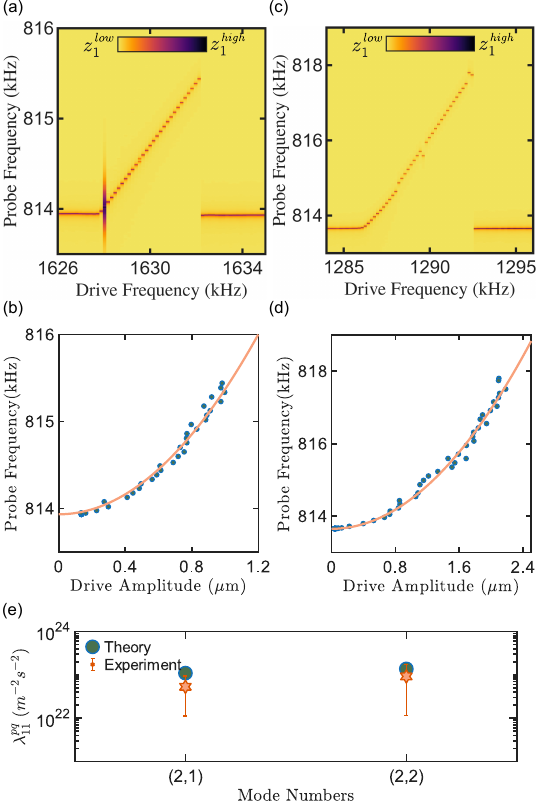}
    \caption{Nonlinear intermodal coupling. (a,c) Resonance frequency-shifting phenomena of the fundamental $(1,1)$ mode induced by strong excitation of the $(2,2)$ and $(2,1)$ modes, respectively.(b,d) Extracted resonance frequency of the $(1,1)$ mode as a function of the displacement amplitude of the driven $(2,2)$ and $(2,1)$ modes, respectively. Solid lines represent fits to the data using Eq. (\ref{9}) to extract the nonlinear intermodal coupling constants. (e) Comparison of experimentally extracted coupling constants (stars) with theoretical values (circles) calculated from Eq. S40.}
    \label{fig:figure 3}
\end{figure}
%

When a higher order mode is driven into the nonlinear regime, its large amplitude oscillation modifies the tensile stress of the membrane. This displacement-dependent tension acts as an effective parametric perturbation for other modes, leading to nonlinear intermodal coupling. As a result the resonance frequency of the fundamental mode becomes dependent on the oscillation amplitude of the driven higher-order mode. To experimentally probe this effect, we strongly drive the $(2,1)$ and $(2,2)$ modes near their resonance frequencies while simultaneously monitoring the response of the fundamental $(1,1)$ mode using a weak probe tone. Figures \ref{fig:figure 3}(a) and \ref{fig:figure 3}(c) show color plots of the fundamental-mode response as a function of probe frequency and drive frequency applied to the higher-order modes. A clear amplitude-dependent frequency shift of the $(1,1)$ mode is observed, indicative of nonlinear frequency pulling due to intermodal coupling. The localized enhancement in spectral amplitude of the fundamental mode observed near a drive frequency $\sim 1628~\mathrm{kHz}$ in Fig. \ref{fig:figure 3}(a) arises from a subharmonic component of the drive signal. This feature is absent in Fig. \ref{fig:figure 3}(c), where such harmonic components are not resonant with the fundamental mode.

In order to theoretically evaluate the intermodal coupling constant, we modify the in-plane stress component as
\begin{equation}
\begin{aligned}
    N_{xx} &= Eh\left[\epsilon_0 \,+\, \frac{1}{2}\left\{\left(\frac{\partial w_1}{\partial x}\right)^2 \,+\, \left(\frac{\partial w_2}{\partial x}\right)^2\right\}\right], \\
    N_{yy} &= Eh\left[\epsilon_0 \,+\, \frac{1}{2}\left\{\left(\frac{\partial w_1}{\partial y}\right)^2 \,+\, \left(\frac{\partial w_2}{\partial y}\right)^2\right\}\right], \\
    N_{xy} &= \frac{1}{2}Eh\left[\left(\frac{\partial w_1}{\partial x}\frac{\partial w_1}{\partial y}\right) \,+\, \left(\frac{\partial w_2}{\partial x}\frac{\partial w_2}{\partial y}\right)\right],
\end{aligned}
\label{7}
\end{equation}
where $w_1 = sin(\frac{n \pi x}{L})\, sin (\frac{m \pi y}{L})\,z_1(t)$ and $w_2 = sin(\frac{p \pi x}{L})\, sin (\frac{q \pi y}{L})\,z_2(t)$ represent the transverse displacements of the probing and driving modes, respectively. Substituting the mode shapes into the governing plate equation and applying the Galerkin discretization procedure, followed by the inclusion of linear damping ($\nu_1$) and external probing force ($f_p \, cos(\omega_p t)$) terms, yields the probe mode  equation of motion as
\begin{equation}
    \begin{aligned}
    \ddot{z_1} &+\omega_{nm}^2 z_1 + \nu_1 \dot{z_1}+\alpha_{nm} z_1^3 + \lambda_{nm}^{pq} z_1 z_{2}^2  = f_p\,  cos(\omega_p t),
    \label{8}
    \end{aligned}
\end{equation}
where $\lambda_{nm}^{pq}$ represents the coupling constant between the driving mode $(p,q)$ and the probing mode $(n,m)$.
%
%
%
Applying the method of multiple scales to Eq. (\ref{8}), we obtain the effective resonance frequency of the fundamental mode arising from intermodal coupling as
\begin{equation}
    \omega = \omega_{nm} + \frac{3\alpha_{nm}}{8 \omega_{nm}} ({z_1^{peak}})^2 + \frac{\lambda_{nm}^{pq}}{2\omega_{nm}}z_{2}^2,
    \label{9}
\end{equation}
where $z_{2}$ is the amplitude of the driven higher order mode. The experimentally measured frequency shifts are shown in Figs. \ref{fig:figure 3}(b) and \ref{fig:figure 3}(d). The data are fitted using Eq. (\ref{9}), with $\lambda_{nm}^{pq}$ as the sole fitting parameter. From these fits, we extract the intermodal coupling constants for the $(1,1)$–$(2,1)$ and $(1,1)$–$(2,2)$ mode pairs, yielding values of $5.3\times10^{22}$ $m^{-2}s^{-2}$and $9.2\times10^{22}$ $m^{-2}s^{-2}$, respectively. As shown in Fig. \ref{fig:figure 3}(e), the extracted coupling constants are in good agreement with the theoretical predictions obtained from Eq. S40. At very high drive amplitudes, the model breaks down due to the onset of higher-order nonlinearities in the driven mode, which modify the standard Duffing response\cite{arora2024mixed,samanta2018tuning}. 

Using the Kirchhoff-Love plate theory, we further compute the nonlinear coupling constants between the fundamental mode and higher-order modes using Eq. S40.%
\begin{figure}
    \centering
    \includegraphics[width=1\linewidth]{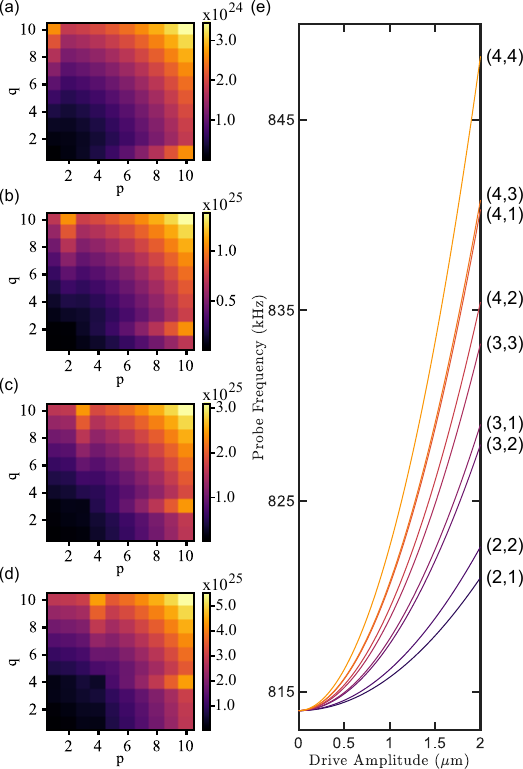}
    \caption{Theoretical estimation of the nonlinear mode coupling matrix. (a–d) Calculated coupling constants as a function of the driven mode indices $(p,q)$ for $p,q = 1$–$10$, coupled to the probe modes $(1,1)$, $(2,2)$, $(3,3)$, and $(4,4)$, respectively. (e) Predicted frequency shift of the fundamental mode as a function of the coupled-mode amplitude, calculated using the theoretical coupling constants listed in Table \ref{tab:Table 1}.}
    \label{fig:Figure 4}
\end{figure}
%
%
We first consider the probe mode with $n=m$. Figure~\ref{fig:Figure 4}(a--d) shows the variation of the coupling constant as the driven mode indices $(p,q)$ are varied from 1 to 10, while the probe modes are fixed at $(1,1)$, $(2,2)$, $(3,3)$, and $(4,4)$, respectively. For driven modes with $p,q \neq n$, the coupling constant $\lambda_{nm}^{pq}$ increases monotonically with increasing $p$ or $q$ when the other index is held fixed. In the special case $p=n$ or $q=m$, $\lambda_{nm}^{pq}$ exhibits a sharp increase with increasing $p$ or $q$, respectively, due to enhanced mode overlap. When $n=m=p=q$, $\lambda_{nm}^{pq}$ reduces to intramodal coupling and equals the Duffing constant of the mode. For probe modes with $n \neq m$, the symmetry is broken (see Supplementary Fig.~S1). In this case, $\lambda_{nm}^{pq}$ exhibits a sharp increase when either $p=n$ or $q=m$. When $n<m$, the coupling constant $\lambda_{nm}^{pq}$ is larger for $p=n$ than for $q=m$. Conversely, when $m<n$, $\lambda_{nm}^{pq}$ is larger for $q=m$ than for $p=n$. When $p=n$ and $q=m$ then $\lambda_{nm}^{pq}$ again reduces to the intramodal coupling constant. The calculated coupling parameters for mode indices $p,q = 1$–$4$, shown in Fig. \ref{fig:Figure 4}(a), are listed in Table \ref{tab:Table 1}. The coupling matrix exhibits clear symmetry with respect to the mode indices and reveals a systematic increase in coupling strength for higher-order modes, as discussed above. Using the coupling parameters in Table \ref{tab:Table 1}, we predict the frequency shift of the fundamental mode as a function of the coupled-mode amplitude, as shown in Fig. \ref{fig:Figure 4}(e). These results highlight nonlinear intermodal coupling as an effective mechanism for frequency control.
\begin{table}
\caption{This table shows the value of coupling constant per unit modal mass for different modes with mode number (p,q), which is coupled to (1,1) mode. All the values here are of the order $10^{23}$\,$m^{-2}s^{-2}$. The intramode coupling constant (Duffing constant) is highlighted in blue.}
\label{tab:Table 1}
\begin{ruledtabular}
\begin{tabular}{ccccc}
        & $q=1$  & $q=2$  & $q=3$  & $q=4$   \\
\hline
$p=1$   & \textcolor{blue}{0.17}   & 1.12   & 2.41   & 4.22    \\
$p=2$   & 1.12   & 1.38   & 2.24   & 3.44    \\
$p=3$   & 2.41   & 2.24   & 3.09   & 4.30    \\
$p=4$   & 4.22   & 3.44   & 4.30   & 5.51    \\
\end{tabular}
\end{ruledtabular}
\end{table}
%

In conclusion, we experimentally demonstrate and quantitatively model nonlinear intra- and intermodal coupling in a high-stress silicon nitride membrane resonator arising from tension induced geometric nonlinearity. By independently driving selected flexural modes into the nonlinear regime, we directly measure amplitude-dependent frequency shifts originating from intrinsic Duffing nonlinearity as well as from nonlinear coupling between orthogonal modes.
Using a Kirchhoff–Love plate theory framework, we derived expressions for the Duffing and intermodal coupling coefficients and showed quantitative agreement with experimental measurements for multiple mode pairs, including the (1,1)-(2,1) and (1,1)-(2,2) interactions. Beyond individual mode pairs, we computed the full nonlinear coupling matrix across mode families, explicitly demonstrating how mode symmetry and spatial overlap determine the magnitude and selectivity of intermodal interactions.
These results establish nonlinear mode coupling in $Si_3N_4$ membranes as a predictable and controllable mechanism for frequency tuning and multimode transduction. The presented framework provides a direct route to engineering mechanical responses in membrane resonators, with immediate implications for multimode sensing, mechanically mediated signal processing, and hybrid opto- and electro-mechanical systems where controlled mode interactions are essential\cite{nielsen2017multimode,mestre2025network,meng2022measurement,zhao2025quantum}.

\bibliographystyle{unsrt}
\bibliography{References}

\begin{center}
\hypertarget{equal}{$^{*}$} These authors contributed equally to this work.\\
\hypertarget{corrA}{$^{\dagger}$} Corresponding author: nishta@iisc.ac.in\\
\hypertarget{corrB}{$^{\ddagger}$} Corresponding author: csamanta@iiserb.ac.in
\end{center}

\vspace{0.5cm}
{\bf Acknowledgments}

A.N. acknowledges funding support from MHRD, MeitY and DST Nano Mission through NNetRA. N.A. acknowledges fellowship support under Visvesvaraya Ph.D. Scheme, Ministry of Electronics and Information Technology (MeitY), India. C.S. acknowledges financial support from the Anusandhan National Research Foundation (ANRF), India, under Project No.~ANRF/ECRG/2024/006637/PMS.

\vspace{0.5cm}

{\bf Competing Interests}

The authors declare no competing interests.

\end{document}


\maketitle

\tableofcontents

\section{Theoretical description}
We model nonlinear mode coupling behavior of a square silicon nitride ($Si_3N_4$) membrane resonator using Kirchhoff–Love plate theory. We first analyze the intramode coupling (intrinsic Duffing nonlinearity) of the resonator and then introduce intermodal mode-coupling terms to develop a general coupling framework. The Duffing constants and a general expression for the coupling constant between probing mode $(n,m)$ and driving mode $(p,q)$ are derived. The theoretical results show good agreement with experimental observations.
\subsection{Duffing nonlinearity}
With the use of Kirchhoff-Love plate theory\cite{leissa1969vibration}, the general equation characterizing the transverse displacement $w(x,y,t)$ of a thin plate can be written as
\begin{align}
    D \nabla^4 w(x, y, t) 
    &- \left( N_{xx} \frac{\partial^2 w}{\partial x^2} 
    + 2N_{xy} \frac{\partial^2 w}{\partial x \partial y} 
    + N_{yy} \frac{\partial^2 w}{\partial y^2} \right)
    + \rho h \frac{\partial^2 w}{\partial t^2} 
    = 0
    \label{S1}
\end{align}
where $w(x,y,t)= \phi(x,y)\, z(t)=\,sin(\frac{n\pi x}{L})\,sin(\frac{m\pi y}{L} )\,z(t)$, D is the flextural rigidity, $N_{xx}$, $N_{yy}$, and $N_{xy}$ are the in-plane stress resultants, which can be defined as

\begin{align}
    N_{xx} &= \int_{-h/2}^{h/2} \sigma_{xx} \, dz, 
    \label{S2}\\
    N_{yy} &= \int_{-h/2}^{h/2} \sigma_{yy} \, dz, 
    \label{S3}\\
    N_{xy} &= \int_{-h/2}^{h/2} \sigma_{xy} \, dz,
    \label{S4}
\end{align}
%
%
%
here, $\sigma_{xx}$, $\sigma_{yy}$ and $\sigma_{xy}$ is defined as
\begin{align}
    \sigma_{xx} &= \frac{E}{1 - \nu^2} (\epsilon_{xx} + \nu \epsilon_{yy}), 
    \label{S5}\\
    \sigma_{yy} &= \frac{E}{1 - \nu^2} (\epsilon_{yy} + \nu \epsilon_{xx}), 
    \label{S6}\\
    \sigma_{xy} &= \frac{E}{1 + \nu} \epsilon_{xy}.
    \label{S7}
\end{align}
%
Due to the thinness of our membrane, we assume that the Poisson's ratio ($\nu$) is zero. In the presence of nonlinear displacement, particularly due to geometrical nonlinearity, the strain components are defined as \cite{ventsel2002thin}
\begin{align}
    \epsilon_{xx} &= \epsilon_0+\frac{\partial u_0}{\partial x} 
    + \frac{1}{2} \left( \frac{\partial w}{\partial x} \right)^2 
    - z(t) \frac{\partial^2 w}{\partial x^2}, 
    \label{S8}\\
    \epsilon_{yy} &= \epsilon_0+\frac{\partial v_0}{\partial y} 
    + \frac{1}{2} \left( \frac{\partial w}{\partial y} \right)^2 
    - z(t) \frac{\partial^2 w}{\partial y^2}, 
    \label{S9}\\
    \epsilon_{xy} &= \frac{1}{2} \left( \frac{\partial u_0}{\partial y} + \frac{\partial v_0}{\partial x} \right) 
    + \frac{1}{2} \frac{\partial w}{\partial x} \frac{\partial w}{\partial y} 
    - z(t) \frac{\partial^2 w}{\partial x \partial y}.
    \label{S10}
\end{align}
For a thin membrane, we assume that in-plane mid-surface displacements are minimal, namely $u_0, v_0 = 0$, and that in-plane displacements resulting from transverse deformation are insignificant. With these assumptions, the stress resultants are simplified to
\begin{align}
    N_{xx} &= Eh \left( \epsilon_0 + \frac{1}{2} \left( \frac{\partial w}{\partial x} \right)^2 \right), \label{S11}\\
    N_{yy} &= Eh \left( \epsilon_0 + \frac{1}{2} \left( \frac{\partial w}{\partial y} \right)^2 \right), \label{S12}\\
    N_{xy} &= \frac{1}{2} Eh \left( \frac{\partial w}{\partial x} \frac{\partial w}{\partial y} \right).
    \label{S13}
\end{align}
Substituting into the governing equation \eqref{S1} and employing the Galerkin method with $w(x,y,t) =  \,sin(\frac{n\pi x}{L})\,sin(\frac{m\pi y}{L} )\,z(t)$, also neglecting $D$ for the high-stress scenario \cite{cattiaux2020geometrical}, we obtain
\begin{align}
    & - \int_0^L \int_0^L 
    \left( N_{xx} \frac{\partial^2 w}{\partial x^2} 
    + 2N_{xy} \frac{\partial^2 w}{\partial x \partial y} 
    + N_{yy} \frac{\partial^2 w}{\partial y^2} \right) 
     \phi(x,y) \, dx \, dy \nonumber \\
    & \quad + \rho h \int_0^L \int_0^L 
    \frac{\partial^2 w}{\partial t^2} 
     \phi(x,y) \, dx \, dy 
    = 0
    \label{S14}
\end{align}
The expressions for $N_{xx}$, $N_{yy}$, and $N_{xy}$  becomes
\begin{align}
    N_{xx} &= Eh \left[ \epsilon_0 + \frac{1}{2} z^2(t) 
    \left( \frac{n\pi}{L} \right)^2 
    \sin^2\left( \frac{m\pi y}{L} \right) \right. \notag \\
    & \quad \left. \times \cos^2\left( \frac{n\pi x}{L} \right) \right], \label{S15}\\
    N_{yy} &= Eh \left[ \epsilon_0 + \frac{1}{2} z^2(t) 
    \left( \frac{m\pi}{L} \right)^2 
    \sin^2\left( \frac{n\pi x}{L} \right) \right. \notag \\
    & \quad \left. \times \cos^2\left( \frac{m\pi y}{L} \right) \right], \label{S16}\\
    N_{xy} &= \frac{1}{2} Eh z^2(t) 
    \left( \frac{n\pi}{L} \right) \left( \frac{m\pi}{L} \right) \notag \\
    & \quad \times \sin\left( \frac{n\pi x}{L} \right) 
    \cos\left( \frac{n\pi x}{L} \right) 
    \sin\left( \frac{m\pi y}{L} \right) 
    \cos\left( \frac{m\pi y}{L} \right).
    \label{S17}
\end{align}
Upon substituting into equation \eqref{S14}, we can find the integrals 
\begin{equation}
    K_1 = \int_0^L \int_0^L N_{xx} \, \frac{\partial^2w}{\partial x^2}\, \, \phi(x,y) \, dx dy
    \label{S18}
\end{equation}
\begin{equation}
    K_2 = \int_0^L \int_0^L N_{yy} \, \frac{\partial^2w}{\partial y^2}\, \, \phi(x,y) \, dx dy
    \label{S19}
\end{equation}
\begin{equation}
    K_3 = \int_0^L \int_0^L 2N_{xy} \, \frac{\partial^2w}{\partial x \partial y}\, \, \phi(x,y) \, dx dy
    \label{S20}
\end{equation}
By solving the integral and substituting these values into Eq. \eqref{S14}, we have derived the equation 
\begin{align}
\rho h\frac{L^2}{4}\frac{d^2z}{dt^2}
&+ Eh\epsilon_0\frac{L^2}{4}\left(\frac{\pi}{L}\right)^2(n^2+m^2)z(t) 
+ \frac{3}{2}Eh\frac{L^2}{64}\left(\frac{\pi}{L}\right)^4
\left[n^4+m^4-\frac{2}{3}n^2m^2\right]z^3(t)
= 0
\label{S21}
\end{align}
From there we can find,
\begin{equation}
   \omega_{nm}^2= \frac{E\epsilon_0\pi^2 (n^2+m^2)}{\rho  L^2}\label{S22}
\end{equation}
And,
\begin{equation}
\alpha_{nm}= \frac{3E}{32\rho}\left(\frac{\pi}{L}\right)^4\left[n^4+m^4-\frac{2}{3}n^2m^2\right]  \label{S23}  
\end{equation}
To determine the resonant frequency and the Duffing nonlinearity, we employed Eq.\eqref{S21}; however, a damping component and a driving force term will typically be present as well. To analyze the nonlinear response, it is essential to include the driving force term, as outlined in the paper. The equation of motion, incorporating both the driving force and damping terms, can be expressed as 
\begin{equation}
    \frac{d^2z}{dt^2} + \nu \frac{dz}{dt} + \omega_{nm}^2 z + \alpha_{nm} z^3 = f_d \cos(\omega t)
    \label{S24}
\end{equation}
here, $\nu$ represents the damping factor, while $f_d \cos(\omega t)$ denotes the driving force. The driving force is either high or low, as indicated by the driving force amplitude $f_d$.

\subsection{Nonlinear mode coupling} 
We have not previously looked at the mode coupling term. The in-plane tension can be adjusted for mode coupling. Considering non-linear mode coupling, the values of $\epsilon_{xx},\, \epsilon_{yy},\, \epsilon_{xy}$ will be 
\begin{equation}
    \epsilon_{xx} = \frac{\partial u_0}{\partial x}\,+\, \frac{1}{2}\left[\left(\frac{\partial w_1}{\partial x}\right)^2\, +\, \left(\frac{\partial w_2}{\partial x}\right)^2\right] - z_1(t)\frac{\partial^2 w_1}{\partial x^2} \label{S25}
\end{equation}
\begin{equation}
    \epsilon_{yy} = \frac{\partial v_0}{\partial y}\,+\, \frac{1}{2}\left[\left(\frac{\partial w_1}{\partial y}\right)^2\, +\, \left(\frac{\partial w_2}{\partial y}\right)^2\right] - z_1(t)\frac{\partial^2 w_1}{\partial y^2} \label{S26}
\end{equation}
\begin{equation}
    \epsilon_{xy} = \frac{1}{2}\left(\frac{\partial u_0}{\partial y}\,+\,\frac{\partial v_0}{\partial x} \right)\, +\, \frac{1}{2}\left[\left(\frac{\partial w_1}{\partial x}\, \frac{\partial w_1}{\partial y}\right)+\left(\frac{\partial w_2}{\partial x}\, \frac{\partial w_2}{\partial y}\right)\right]-z_1(t)\frac{\partial^2\omega_1}{\partial x\partial y} \label{S27}
\end{equation}
Here, \( w_1 = \phi_1(x,y)z_1(t) = \sin\left(\frac{n\pi x}{L}\right) \sin\left(\frac{m\pi y}{L}\right) z_1(t) \), where \( (n, m) \) denotes the mode number of the probe mode, and \( w_2 = \phi_2(x,y)z_2(t)= \sin\left(\frac{p\pi x}{L}\right) \sin\left(\frac{q\pi y}{L}\right) z_2(t) \), where \( (p, q) \) signifies the mode number corresponding to the pump mode. Once more, we can apply the aforementioned thin membrane approximation. The in-plane stress resultants can now be expressed as 
\begin{equation}
    N_{xx}= Eh\left[\epsilon_0 \, + \frac{1}{2}\left\{\left(\frac{\partial w_1}{\partial x}\right)^2\, +\, \left(\frac{\partial w_2}{\partial x}\right)^2\right\}\right]
    \label{S28}
\end{equation}
\begin{equation}
    N_{yy}= Eh\left[\epsilon_0 \, + \frac{1}{2}\left\{\left(\frac{\partial w_1}{\partial y}\right)^2\, +\, \left(\frac{\partial w_2}{\partial y}\right)^2\right\}\right]
    \label{S29}
\end{equation}
\begin{equation}
    N_{xy} = \frac{1}{2}Eh\left[\left(\frac{\partial w_1}{\partial x}\frac{\partial w_1}{\partial y}\right)\,+\,\left(\frac{\partial w_2}{\partial x}\frac{\partial w_2}{\partial y}\right)\right]
    \label{S30}
\end{equation}
The contributions to the mode coupling factor from the terms $N_{xx},\, N_{yy},\, N_{xy}$ are as follows 
\begin{equation}
    N_{xx,\lambda} = \frac{Eh}{2}\left(\frac{\partial w_2}{\partial x}\right)^2
    \label{S31}
\end{equation}
\begin{equation}
    N_{yy,\lambda} = \frac{Eh}{2}\left(\frac{\partial w_2}{\partial y}\right)^2
    \label{S32}
\end{equation}
\begin{equation}
    N_{xy,\lambda} = \frac{Eh}{2}\left(\frac{\partial w_2}{\partial x}\frac{\partial w_2}{\partial y}\right)
    \label{S33}
\end{equation}
Substituting the values of $w_2$ yields 
\begin{equation}
    N_{xx,\lambda} = \frac{Eh}{2}\left[ \,\left(\frac{p \pi}{L}\right)^2\, cos^2\left(\frac{p\pi x}{L}\right)\, sin^2\left(\frac{q\pi y}{L}\right)\right]
    \label{S34}
\end{equation}
\begin{equation}
    N_{yy,\lambda} = \frac{Eh}{2}\left[ \,\left(\frac{q \pi}{L}\right)^2\, sin^2\left(\frac{p\pi x}{L}\right)\, cos^2\left(\frac{q\pi y}{L}\right)\right]
    \label{S35}
\end{equation}
\begin{equation}
    N_{xy, \lambda} = \frac{1}{2}Eh\left[ \left(\frac{p\pi}{L}\, \right)\, cos\left(\frac{p\pi x}{L}\right)\, sin\left(\frac{q\pi y}{L}\right) \left(\frac{q\pi}{L}\, \right)\, sin\left(\frac{p\pi x}{L}\right)\, cos\left(\frac{q\pi y}{L}\right)\right]
    \label{S36}
\end{equation}
To calculate the coupling constant, it is necessary to solve the integrals 
\begin{equation}
    \int_0^L \int_0^L N_{xx.\lambda}\frac{\partial^2 w_1}{\partial x^2}\, \, \phi_1(x,y) \,dx dy \label{S37}
\end{equation}
\begin{equation}
    \int_0^L \int_0^L N_{yy.\lambda}\frac{\partial^2 w_1}{\partial y^2}\, \, \phi_1(x,y) \,dx dy \label{S38}
\end{equation}
\begin{equation}
    \int_0^L \int_0^L 2N_{xy.\lambda}\frac{\partial^2 w_1}{\partial x\partial y}\, \, \phi_1(x,y) \,dx dy \label{S39}
\end{equation}
The expression for coupling constant is 
\begin{equation}
    \lambda_{nm}^{pq} = \frac{-\int_0^L \int_0^L \left (N_{xx,\lambda}\frac{\partial^2 w_1}{\partial x^2}\, \, \phi_1(x,y) \, + \, N_{yy,\lambda}\frac{\partial^2 w_1}{\partial y^2}\, \, \phi_1(x,y) \, + \, 2N_{xy,\lambda}\frac{\partial^2 w_1}{\partial x\partial y}\, \, \phi_1(x,y)\right) \,dx dy}{\rho h \left(\frac{L^2}{4}\right)}
    \label{S40}
\end{equation}
Where the denominator comes from Eq.\eqref{S21}. Considering the mode number $(p,q)$, where $p=q >1$, when coupled with mode (1,1), the expression for the coupling constant can be defined as 
\begin{equation}
    \lambda_{11}^{pq} = \frac{E}{8\rho}(\frac{\pi}{L})^4[p^2+q^2] \label{S41}
\end{equation}
We have derived an equation similar to \eqref{S24} that includes the mode coupling term. The equation may be expressed as 
\begin{equation}
    \frac{d^2z_1}{dt^2} + \nu_1 \frac{dz_1}{dt} + \omega_{nm}^2 z_1 + \alpha_{nm} z_1^3 + \lambda_{nm}^{pq} z_1 z_2^2 = f_p \cos(\omega t)
    \label{S42}
\end{equation}
This equation represents the mode coupling of the probe mode, which exhibits displacement along the $z$ axis
$(z_1)$, with the driving mode, whose displacement is 
$z_2$. $\nu_1$, $\omega_{nm}$, $\alpha_{nm}$, and $\lambda_{nm}^{pq}$ represent the damping constant, natural frequency, Duffing constant of the probe mode, respectively and $\lambda_{nm}^{pq}$ is the coupling constant between probe mode $(n,m)$ and drive mode $(p,q)$.
%
We have numerically determined the coupling constant using Equation \eqref{S40}. When $p = n$ and $q = m$, the coupling constant $\lambda^{nm}_{pq}$ reduces to $\alpha_{nm}$, corresponding to the intramode coupling, which is basically the Duffing constant.
\\
\begin{table}[H]
\centering
\begin{minipage}{0.48\textwidth}
\centering
\caption{Coupling constant for unit modal mass for mode $(p,q)$, where $p, q$ runs from 1 to 5, coupled with (2,2) mode. The values of the coupling constant are of the order $10^{23}$ $m^{-2}s^{-2}$.The intramode coupling constant (Duffing constant) is highlighted in blue.}
\begin{tabular}{c c c c c c}
\rowcolor{gray!20}
\toprule
& $q = 1$ & $q = 2$ & $q = 3$ & $q = 4$ & $q = 5$ \\
\midrule
$p = 1$ & 1.38 & 2.41 & 6.88 & 11.70 & 17.89 \\
$p = 2$ & 2.41 & \textcolor{blue}{2.75} & 10.67 & 17.89 & 27.19 \\
$p = 3$ & 6.88 & 10.67 & 12.39 & 17.21 & 23.40 \\
$p = 4$ & 11.70 & 17.89 & 17.21 & 22.03 & 28.22 \\
$p = 5$ & 17.89 & 27.19 & 23.40 & 28.22 & 34.42 \\
\bottomrule
\label{Table S1}
\end{tabular}
\end{minipage}\hfill
\begin{minipage}{0.48\textwidth}
\centering
\caption{Coupling constant for unit modal mass for mode $(p,q)$, where $p, q$ runs from 1 to 5, coupled with (3,3) mode. The values of the coupling constant are of the order $10^{23}$ $m^{-2}s^{-2}$.The intramode coupling constant (Duffing constant) is highlighted in blue.}
\begin{tabular}{c c c c c c}
\rowcolor{gray!20}
\toprule
& $q = 1$ & $q = 2$ & $q = 3$ & $q = 4$ & $q = 5$ \\
\midrule
$p = 1$ & 3.09 & 7.74 & 9.29 & 26.33 & 40.27 \\
$p = 2$ & 7.74 & 12.39 & 16.26 & 30.97 & 44.91 \\
$p = 3$ & 9.29 & 16.26 & \textcolor{blue}{13.93}  & 44.14 & 65.05 \\
$p = 4$ & 26.33 & 30.97 & 44.14 & 53.60 & 63.50 \\
$p = 5$ & 40.27 & 44.91 & 65.05 & 63.60 & 77.44 \\
\bottomrule
\label{Table S2}
\end{tabular}
\end{minipage}
\end{table}


\begin{figure}[H]
    \centering
    \includegraphics[width=0.7\linewidth]{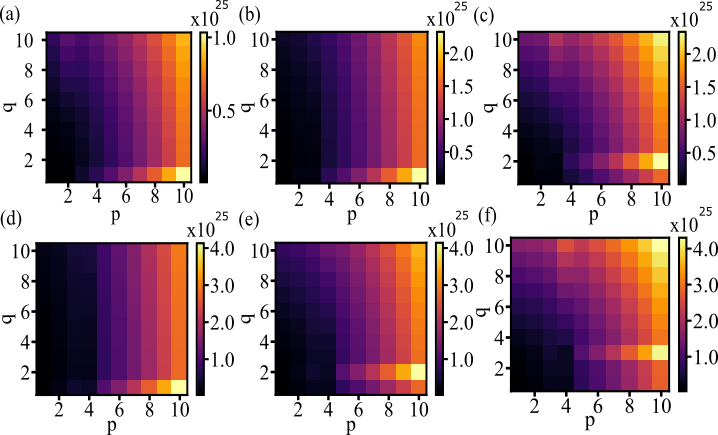}
    \caption{Theoretical estimation of the nonlinear mode coupling matrix. (a–f) Calculated coupling constants as a function of the driven mode indices $(p,q)$ for $p,q = 1$–$10$, coupled to the probe modes $(2,1)$, $(3,1)$, $(3,2)$, $(4,1)$, $(4,2)$ and $(4,3)$, respectively. }
    \label{Figure S1}
\end{figure}

\subsection{Nonlinear frequency pulling}
The frequency detuning equation for mode coupling is derived using the method of multiple scales\cite{kozinsky2007nonlinear}. This approach alters the time scale as follows 
\begin{equation}
    T_n = \epsilon^n t,
    \label{S43}
\end{equation}
where $T_0 = t$, and $\epsilon$ is the scaling parameter. For using method of multiple scale the Eq \eqref{S42} can be modified as
\begin{equation}
    \ddot{z}_1 + 2 \epsilon\mu_1 \dot{z}_1 + \omega_{nm}^2 z_1 + \epsilon \alpha_{nm} z_1^3 + \epsilon \lambda_{nm}^{pq} z_1 z_2^2 = \epsilon k_d \, cos(\omega_{nm}T_0 + \sigma T_1).
    \label{S44}
\end{equation}
Using the method of multiple scale method we have reached a equation as 
\begin{equation}
    \sigma = \frac{3\alpha_{nm}}{8\omega_{nm}}({z_1^{peak}})^2 \pm \sqrt{\frac{k_d^2}{4\omega_{nm}^2({z_1^{peak}})^2}-\mu_1 ^2}\, +\frac{\lambda_{nm}^{pq}}{2\omega_{nm}}z_{2}^2 
    \label{S45}.
\end{equation}
For steady state the middle term in the RHS becomes zero. Then we can get a equation like
\begin{equation}
    \sigma = \frac{3\alpha_{nm}}{8\omega_{nm}}({z_1^{peak}})^2 + \frac{\lambda_{nm}^{pq}}{2\omega_{nm}}z_{2}^2.
    \label{S46}
\end{equation}
Here, $\sigma = \omega - \omega_{nm}$ denotes the frequency detuning. This equation represents the frequency detuning generated by mode coupling, enabling the determination of the coupling constant ($\lambda_{nm}^{pq}$). In the absence of mode coupling, the equation assumes the following form 
\begin{equation}
     \sigma = \frac{3\alpha_{nm}}{8\omega_{nm}}({z_1^{peak}})^2.
     \label{S47}
\end{equation}

\begin{figure}[H]
    \centering
    \includegraphics[width=0.9\linewidth]{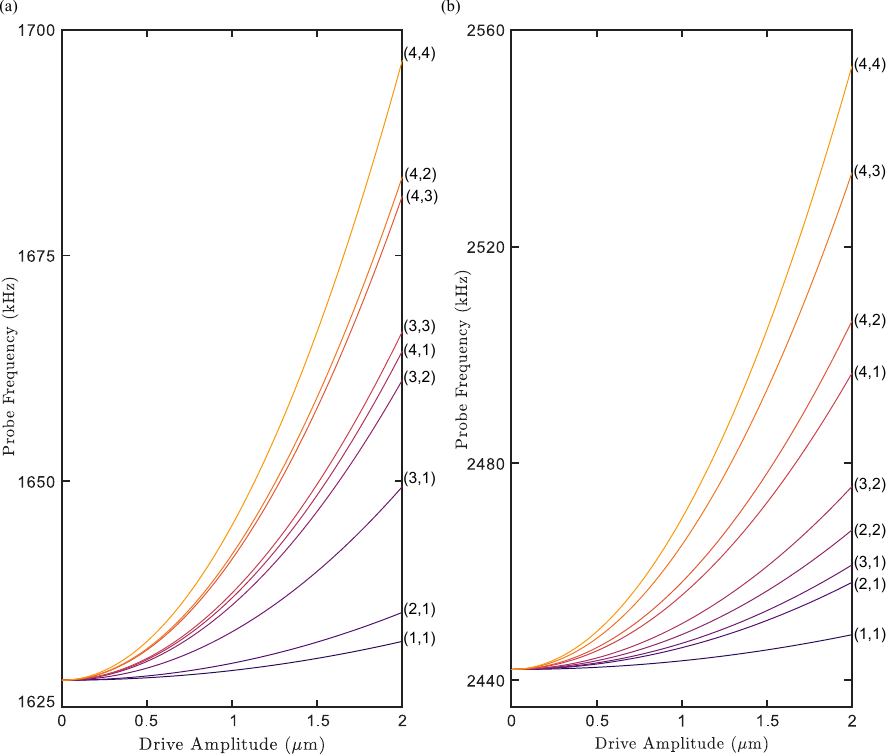}
    \caption{Predicted frequency shift of the fundamental mode (a)(2,2), (b)(3,3) as a function of the coupled-mode amplitude.}
    \label{Figure S2}
\end{figure}

Equation \eqref{S47} helps us for the calculation of the experimental Duffing constant ($\alpha_{nm}$). When the applied force is sufficiently small, the nonlinear terms in Eq. \eqref{S45} can be neglected. Using the coupling constant, the resulting shift in the resonance frequency of the probe mode can then be plotted as a function of the drive-mode amplitude. The resonance-frequency shifts of the (2,2) and (3,3) modes as functions of the peak amplitudes of the other coupled modes are shown in Fig. \ref{Figure S2}. The graph illustrates the extent of the change in resonant frequency of the modes that can occur when the coupled mode possesses the specified amplitude.


\section{Experimental Section}

\subsection{Device parameters}

The device parameters used in this work are summarized in Table~\ref{Table S1}. The strain is calculated using the resonance frequency relation given in Eq.\eqref{S22}. Using the extracted strain and the Young’s modulus, the corresponding tensile stress is estimated to be approximately 1.05~GPa, consistent with previously reported values~\cite{shu2009fabrication}.

\vspace{0.5 cm}
%


\begin{table}[H]
\centering
\begin{tabular}{|p{6cm}|p{6cm}|}
\hline
\textbf{Physical Parameter} & \textbf{Value} \\
\hline
Young's Modulus (E) & $280\times10^9 \,\text{Pa}$ \\
\hline
Density ($\rho$) & $3170 \,\text{kg/m}^3$ \cite{mestre2025network} \\
\hline
Inbuilt Strain ($\epsilon_0$) & $3.751\times 10^{-3}$ \\
\hline
Length (L) & $500\times10^{-6} \,\text{m}$ \\
\hline
Thickness (h) & $100\times 10^{-9} \,\text{m}$ \\
\hline
\end{tabular}
\caption{Device Parameters}
\label{Table S1}
\end{table}

\subsection{Displacement calibration}

\begin{figure}[H]
    \centering    \includegraphics[width=0.5\linewidth]{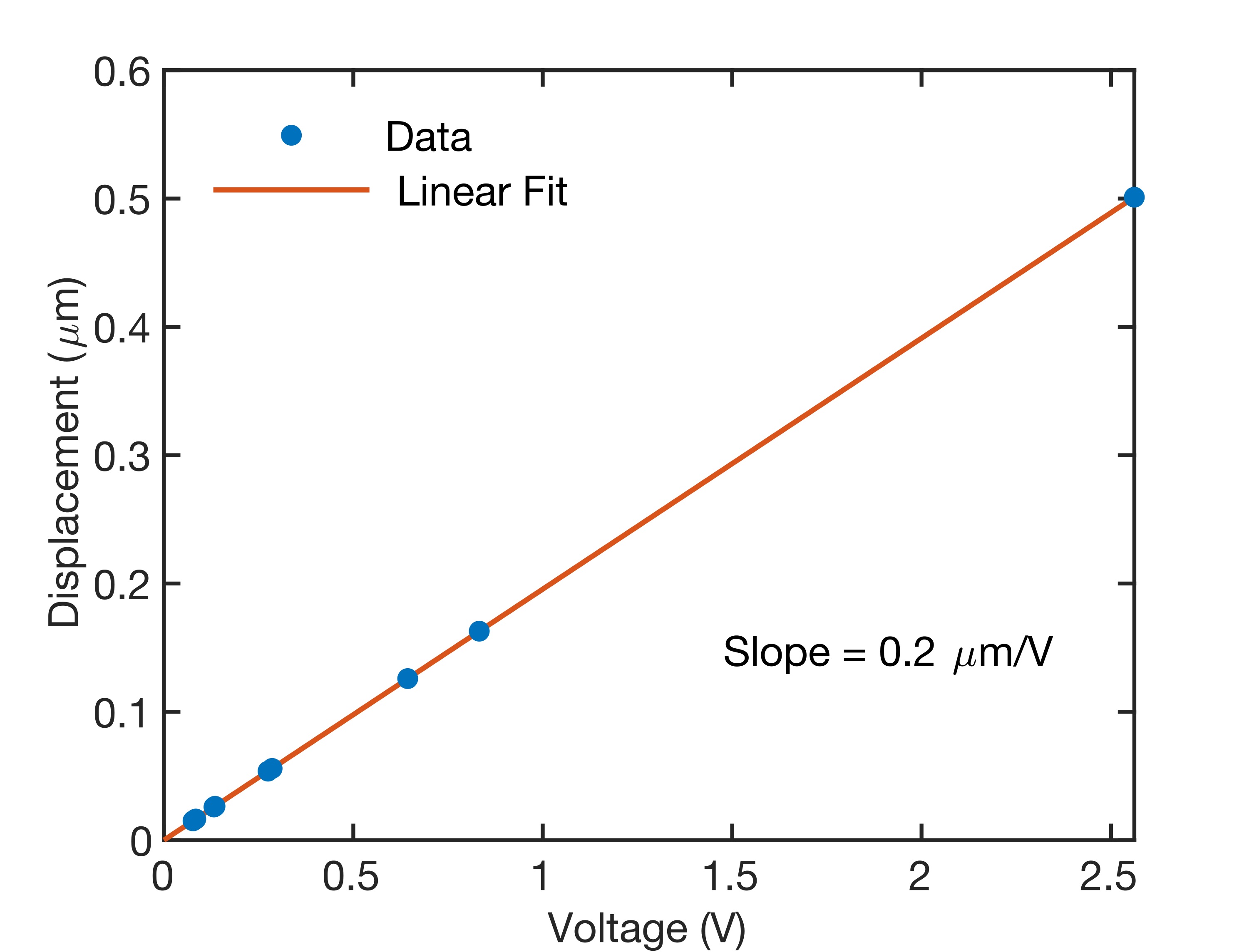}
    \caption{Measured mechanical displacement $z$ as a function of the detected lock-in amplifier voltage $V$. A linear fit to the data yields a displacement calibration factor of $\partial z / \partial V = 0.2~\mu\mathrm{m}\,\mathrm{V}^{-1}$.}
    \label{Figure S3}
\end{figure}

The displacement of the membrane was calibrated using the laser Doppler vibrometer (LDV), which provides an absolute measurement of the out-of-plane displacement. To establish a conversion between the electrical readout voltage and the physical displacement, the device was driven with a low-frequency sinusoidal excitation of known amplitude while simultaneously recording the displacement using the LDV and the corresponding voltage signal using the lock-in amplifier. Figure ~\ref{Figure S3} shows the measured displacement as a function of the detected voltage. A linear fit to the data yields a calibration factor of $\frac{\partial z}{\partial V} = 0.2~\mu\mathrm{m}/\mathrm{V}$, where $z$ is the mechanical displacement and $V$ is the measured voltage at the lock-in amplifier.

This calibration factor was subsequently used to convert all voltage signals measured using the lock-in amplifier into absolute displacement units. Specifically, the displacement amplitude was obtained as $z(t) = \alpha V(t)$, where $\alpha = 0.2~\mu\mathrm{m}/\mathrm{V}$ is the experimentally determined conversion factor. Since the LDV provides an absolute displacement reference traceable to optical wavelength standards, this procedure enables quantitative displacement measurements without reliance on device parameters.

\subsection{Frequency responses of modes $(2,1)$ and $(2,2)$}

\begin{figure}[H]
    \centering
    \includegraphics[width=\linewidth]{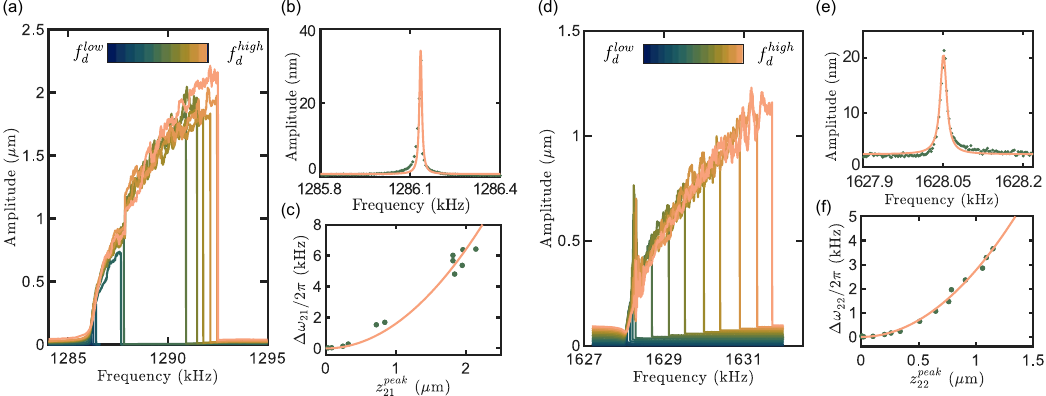}
    \caption{(a,d) Nonlinear frequency responses of mode $(2,1)$ and $(2,2)$ at increasing driving force. (b,e) Linear responses of the $(2,1)$ and $(2,2)$ modes at the lowest drive amplitudes; solid lines indicate Lorentzian fits used to extract the resonance frequencies and quality factors. (c,f) Amplitude-dependent frequency shifts corresponding to the maximum displacement amplitudes extracted from (a,d). Solid lines show fits to extract the Duffing coefficients.}
    \label{Figure S4}
\end{figure}

Figures \ref{Figure S4}(b,e) show the linear frequency responses of the $(2,1)$ and $(2,2)$ modes respectively. The solid lines represent Lorentzian fits to the data, yielding quality factors of \(12.5\times10^{4}\) for the $(2,1)$ mode and \(11.5\times10^{4}\) for the mode $(2,2)$. As the driving force is increased, the frequency response becomes progressively nonlinear, as illustrated in Figs.~\ref{Figure S4}(a,d).  Figures~\ref{Figure S4}(c,f) show the extracted frequency shift as a function of the peak amplitudes. The solid line correspond to fits using Eq.~\eqref{S47}, from which the experimental Duffing coefficients are obtained to be \(2.5\times10^{23}\,\mathrm{m^{-2}s^{-2}}\) and \(4.7\times10^{23}\,\mathrm{m^{-2}s^{-2}}\) for mode $(2,1)$ and $(2,2)$ respectively.

Throughout the measurements, the LDV laser spot was fixed at the center of the membrane. This position corresponds to the anti-node of the (1,1) mode but lies close to the nodal regions of the higher-order (2,1) and (2,2) modes. As a result, the measured displacement amplitudes for these modes are small, leading to increased amplitude fluctuations, as observed in Fig.~\ref{Figure S4}(a,d). To account for this effect, scaling factors were estimated from the extrema of the spatial mode shapes shown in Fig. 1 in the main text. The scaling factor is $10.2 \pm 4.0$ for the (2,1) mode and $25.2 \pm 11.0$ for the (2,2) mode. The uncertainty arises from the mismatch between the LDV laser spot size ($\sim 1~\mu\mathrm{m}$) and the step size used in the mode-shape measurements ($\sim 50~\mu\mathrm{m}$). In addition, the localized peak near $\sim 1628~\mathrm{kHz}$ is attributed to higher-harmonic contributions from the parametrically driven fundamental mode.

%



\printbibliography